\documentclass[conference]{IEEEtran}
\IEEEoverridecommandlockouts

\usepackage{blindtext}
\usepackage{cite}
\usepackage{amsmath,amssymb,amsfonts}
\usepackage{enumitem}
\usepackage[linesnumbered,ruled,vlined]{algorithm2e}
\usepackage[noend]{algpseudocode}
\usepackage{graphicx}
\usepackage{textcomp}
\usepackage{svg}
\usepackage{xcolor}
\usepackage[utf8]{inputenc}
\usepackage{lipsum}

\def\BibTeX{{\rm B\kern-.05em{\sc i\kern-.025em b}\kern-.08em
    T\kern-.1667em\lower.7ex\hbox{E}\kern-.125emX}}
\usepackage[left= 1.6 cm, right= 1.6 cm, top=1.778 cm, bottom = 2.65cm]{geometry}
\begin{document}

\title{Hierarchical Deep Reinforcement Learning for Robust Access in Cognitive IoT Networks under Smart Jamming Attacks}


\author{\IEEEauthorblockN{Nadia Abdolkhani and Walaa Hamouda}
\IEEEauthorblockA{Department of Electrical and Computer Engineering \\
Concordia University,
Montreal,  Quebec, H3G 1M8, Canada\\
email:\{n\_abdolk, hamouda\}@ece.concordia.ca}}
\maketitle

\begin{abstract}
In this paper, we address the challenge of dynamic spectrum access in a cognitive Internet of Things (CIoT) network where a secondary user (SU) operates under both energy constraints and adversarial interference from a smart jammer. The SU coexists with primary users (PUs) and must ensure that its transmissions do not exceed a predefined interference threshold on licensed channels. At each time slot, the SU must jointly determine whether to transmit or harvest energy, which channel to access, and the appropriate transmit power while satisfying energy and interference constraints. Meanwhile, a smart jammer actively selects a channel to disrupt, aiming to degrade the SU’s communication performance. This setting presents a significant challenge due to its multi-level decision structure and hybrid action space, which combines both discrete and continuous decisions. To tackle this, we propose a novel Hierarchical Deep Deterministic Policy Gradient (H-DDPG) framework that decomposes the decision-making process into three levels: the high-level policy determines the mode (transmit or harvest), the mid-level policy selects the channel, and the low-level actor outputs a continuous power level. Concurrently, the jammer is modeled as a reinforcement learning agent that learns an adaptive channel jamming strategy using a discrete variant of DDPG. Simulation results show that our H-DDPG approach outperforms conventional flat reinforcement learning baselines.
\end{abstract} 
\begin{IEEEkeywords}
cognitive radio networks, spectrum sharing, smart jammer, deep reinforcement learning.
\end{IEEEkeywords}
\section{Introduction}
\IEEEPARstart{T}{he} exponential growth of wireless devices and Internet of Things (IoT) has led to severe spectrum scarcity, making efficient spectrum management imperative. Cognitive IoT (CIoT) networks have emerged as a promising solution by enabling secondary users (SUs) to opportunistically access underutilized licensed channels \cite{nada_survey_2023}. However, practical CIoT systems face critical challenges, such as energy constraints and adversarial interference from jammers \cite{10700946}. There are three principal CIoT access paradigms enabling spectrum utilization by SUs: interweave, underlay, and overlay. In the interweave model, SUs are permitted to transmit only when the spectrum is idle and primary users (PUs) are inactive \cite{10829806}. The underlay model allows concurrent transmission by SUs and PUs, provided that the interference from SUs remains below a tolerable threshold \cite{10937228}. Lastly, the overlay approach permits SUs to access the spectrum by aiding PUs, typically through cooperative techniques such as relaying or coding, in return for access privileges \cite{10572320}.

Traditional spectrum access techniques often assume static environments and fail to capture the dynamic nature of real-world decisions, such as access strategy, channel selection, and power control. Effective power control is especially critical for energy-constrained devices, which must balance the dual objectives of maximizing data throughput and prolonging network lifetime. Energy harvesting (EH) has emerged as a viable solution to address these energy limitations, enabling devices to become self-sustaining by drawing energy from ambient sources. This not only reduces dependence on external power supplies but also ensures long-term network operability and resilience.

Recent advances in deep reinforcement learning (DRL) have shown promise in enabling autonomous decision-making in dynamic and partially observable wireless environments \cite{Ali_Naser_Muhaidat_2023}. Although CIoT networks benefit from the opportunistic spectrum access, they are vulnerable to jamming attacks given the inherent shared and broadcast nature of radio wave propagation \cite{Lin_Qiu_Wang_Zhang_2024}. However, applying flat DRL models to CIoT systems with complex action spaces often leads to poor convergence and scalability issues. Furthermore, in the presence of intelligent adversaries like smart jammers, the SU must dynamically adapt its behavior across multiple dimensions while preserving long-term performance. In most studies, the action-space is simplified, ignoring the multi-level nature and the hybrid nature of the problem. Authors in \cite{Nadia_Jamming_IoTJ_2024} developed a double deep Q-network (DDQN) designed to help a CIoT agent learn the optimal communication policy navigating a random jammer, under the assumptions of a single channel and a fully discrete action space. \cite{DDQN_jamm_channelSelection} optimized the channel selection problem using the DDQN algorithm to counter the jamming attacks and maximize the successful transmission of the agents. Authors in \cite{Qlearning_antiJam} considered a markov game-based Q-learning approach to channel selection strategy for mitigating the intelligent jammer attacks. Furthermore, \cite{AC_antiJamm} proposed a transfer game actor-critic framework to optimize channel selection and improve communication security against jamming attacks.

To the best of the authors' knowledge, no previous research has considered designing a hierarchical learning algorithm that aims to maximize the CIoT network's throughput under interference limits, energy constraints, and smart jamming attacks. The main contributions of the paper are summarized as follows:

\begin{itemize}
    \item Hierarchical Learning Framework: We propose a Hierarchical Deep Deterministic Policy Gradient (H-DDPG) framework tailored for multi-level, hybrid decision-making in energy-constrained underlay CIoT networks operating in the presence of a smart jammer.

    \item Multi-Level Decision Decomposition: The decision process of the SU is decomposed into three sequential stages: (1) mode selection (transmit or harvest), (2) channel selection, and (3) transmit power control.

    \item Hybrid Action Representation: Our hierarchical approach effectively captures the hybrid nature of the action space by handling discrete and continuous actions separately. This contrasts with prior works that simplify the problem to only channel selection or only power allocation.

    \item Realistic Energy Modeling: We incorporate realistic energy constraints by modeling battery-powered SU devices, in contrast to existing works that assume idealized or unlimited energy availability.

    \item Scalability and Efficiency: The hierarchical decomposition alleviates the action-space explosion problem commonly encountered in DRL, thereby improving scalability and reducing computational and energy overhead.

    \item Adaptive Jammer Modeling: A smart jammer is modeled using a dedicated DDPG agent that learns an optimal jamming strategy through environment interaction, outperforming static and random jamming methods and providing a more realistic adversary for the CIoT system.

    \item We offer a comprehensive analysis of convergence and performance of the proposed algorithm that is benchmarked against alternative methodologies found in existing literature across various test scenarios.
\end{itemize}

The rest of the paper is structured as: Section II presents the system model and problem formulation, Section III showcases the smart jammer model, Section IV presents the proposed H-DDPG framework, Section V discusses the simulation model and presents a comprehensive analysis of the paper's results, and finally, Section VI concludes the paper.
\section{System Model}

Consider a time-slotted underlay CIoT system with N SUs, M PUs, and J smart jammers. Each PU $m$, $1\leq m\leq M$, is associated with a dedicated licensed channel, exclusively occupying its assigned channel for a maximum of $L_m$ time slots during a transmission frame of length $T$ slots, $1<L_m<T$. Each PU $m$ can consistently transmit at a power level of $P_{pm}^t$ at any slot. The time horizon consists of $T$ discrete slots, each of the duration of $\tau$ seconds. The SU $n$, $1\leq n\leq N$, has a finite battery capacity $B_n^{\max}$ and harvests energy from ambient RF signals. 
The SU $n$ possesses the capability
to autonomously and dynamically regulate its transmit power $P_{sn}^t$.

In underlay CIoT, the SU $n$ can use the same time slot as the PU $m$ as long as it adheres to the interference threshold $I_m^{th}$. Therefore, the SU device needs to decide on its transmission power $P_{sn}^t$ such that it adheres to the interference constraint given as
\begin{equation}
    P_{sn}^tg_{nm}^t \leq I_m^{th},
\end{equation} where $g_{nm}^t$ is the channel power gain between the SU $n$ and PU $m$.  The PU status indicator is defined as
\begin{equation}\label{eq:PU_indicator}
  \omega_{pm}^t =
  \begin{cases}
    1 & \text{if the PU $m$ is using time slot $t$}, \\
    0 & \text{otherwise.}
  \end{cases}
\end{equation}
The power gains of the channel between the SU $n$ Tx-Rx pair is $g_{nn}^t$, the PU $m$ and the SU $n$ is $g_{mn}^t$, and the SU $n$ and the PU $m$ is $g_{nm}^t$, are characterized as Rayleigh fading channels that are independently and identically distributed (i.i.d). It is assumed that these channel power gains remain constant within a time slot, but may vary independently from one time slot to another. 

The SU $n$ transmitter is capable of charging its finite battery, denoted by $B_n^{max}$, through RF EH.
Initially, the harvested energy is set to zero. It is assumed that the amount of energy collected in each time slot, $e_n^t$, follows a uniform distribution ranging from $0$ to $E_{max}$, where $e_n^t \sim U(0, E_{max})$. 
It is important to highlight that the energy harvesting process does not lead to any additional energy consumption for PUs or other devices \cite{9606870}.
The initial battery level of SU $n$ is denoted as $B_n^0$, with $B_n^t$ representing the energy available in the battery at the $t$-th time slot. In the next time slot $t+1$, the available energy is updated based on the SU $n$'s decision $d_n^t$ as \vspace{-0.05cm}\begin{equation} \label{eq:eq01}
    B_n^{t+1} = \text{min}\big\{
    B_n^t + d_n^t e_n^t -(1-d_n^t)P_{sn}^t\tau, B_n^{max}\big\},
\end{equation}
where $d_n^t$= 0 indicates that the SU decided to transmit and $d_n^t$= 1 indicates that the SU will harvest energy.
Moreover, any harvested energy that exceeds the battery's maximum capacity is deemed to be discarded. 

When the selected channel is idle (i.e., not occupied by the PU), the SU $n$’s achievable rate at time slot \( t \) is given by:
\begin{equation}
    R_{n0}^t = \log_2\left(1 + \frac{P_{sn}^t g_{nn}^t}{\sigma^2} \right),
\end{equation}
where \( \sigma^2 \) is the noise power.
If the PU $m$ is active on the selected channel, the SU experiences interference from the PU. Consequently, the achievable rate is reduced and given by:
\begin{equation}
    R_{n1}^t = \log_2\left(1 + \frac{P_{sn}^t g_{nn}^t}{P_{pm}^t g_{mn}^t + \sigma^2} \right),
\end{equation}
where \( P_{pm}^t \) and \( g_{mn}^t \) represent the PU's transmit power and the channel gain from the PU transmitter to the SU receiver, respectively.

Taking into account both PU activity and potential jamming, the effective achievable rate of the SU at time slot \( t \) is expressed as:
\begin{equation}
    R_n^t = \omega_{jm}^t (1 - d_n^t) \left[ (1 - \omega_{pm}^t) R_{n0}^t + \omega_{pm}^t R_{n1}^t \right],
    \label{eq:effective_rate}
\end{equation}
where \( \omega_{jm}^t \in \{0, 1\} \) is the jammer’s activity indicator, with \( \omega_{jm}^t = 0 \) indicating a jamming attack effectively reducing the SU achievable rate to zero, and \( \omega_{pm}^t \in \{0, 1\} \) is the PU occupancy indicator. The term \( d_n^t \in \{0, 1\} \) denotes the SU’s access decision, where \( d_n^t = 1 \) indicates energy harvesting mode and \( d_n^t = 0 \) indicates active transmission. 


In the studied model, the SU agent seeks to maximize its rate while considering the interference threshold $I_m^{th}$, the available energy of the battery $B_n^t$, and the energy harvested $e_n^t$. The SU device needs to learn how to optimize both its transmission power \(P_{sn}^t\) and decisions  \(d_n^t\) (mode selection) and \(c_n^t\) (channel selection) to maximize the throughput of the network. Therefore, the maximization of the rate of the SU is modeled as a constrained optimization problem as

\begin{subequations}\small
\label{eqn:optim}
\begin{align}
   &\max_{P_{sn}^t, c_n^t} \sum_{t=1}^{T}\sum_{n=1}^{N} \omega_{jm}^t (1 - d_n^t) \left[ (1 - \omega_{pm}^t) R_{n0}^t + \omega_{pm}^t R_{n1}^t \right], \label{maximization}\\
   &\text{s.t.  } \sum_{t=1}^{k} P_{sn}^t\tau \leq B_n^0+\sum_{t=0}^{k-1}e_n^t, ~~\forall k, \label{constraint1}\\
   & ~~~~~0\leq (1-d_n^t)P_{sn}^t\tau\leq B_n^t,~~ d_n^t\in I\triangleq\{0,1\}, \label{constraint2}\\
   & ~~~~~d_n^t = 1 , ~~ \forall \omega_{jm}^t=0, \label{constraint3}\\
   & ~~~~~\omega_{pm}^tg_{nm}^tP_{sn}^t\leq I_m^{th},~~ \omega_{pm}^t\in\Omega\triangleq\{0,1\} .\label{constraint4}
\end{align}
\end{subequations} 
where $k$ is the number of slots the SU decides to transmit. 

\section{Smart Jammer Model}
We consider the presence of a smart jammer functioning as an adversarial agent within the CIoT environment. The jammer’s primary objective is to disrupt SU transmissions by strategically selecting channels to jam, thereby degrading successful transmission rates and overall system efficiency. We assume that the jammer targets only SU communications, avoiding PUs. This assumption is based on two key considerations: first, the significant legal and technical risks associated with jamming licensed PU transmissions; and second, physical or protocol constraints that limit the jammer’s ability to interfere with PUs. In real-world settings, methods such as cyclostationary detection or matched filtering can help the jammer distinguish between PU and SU signals.

At the beginning of each time slot, the jammer first determines PU activity and launches a jamming attack. The jammer transmits with a power level $P_j^t$ to interfere with the selected channel. Importantly, the SUs have no prior knowledge of which channels will be jammed at any given time slot. To fulfill its objective, the jammer is modeled as an RL agent that observes the global state of the environment and learns a channel selection strategy over time. Specifically, we implement the jammer as a DDPG agent adapted for a discrete action space. At each time step $t$, the jammer observes the system state $s_t$ and selects an action $a_j^t \in {0, 1, \dots, M-1}$, corresponding to the channel index to jam.

The jammer is trained to maximize its long-term impact on SU performance. Its reward at time $t$ is defined as $r_j^t = -R^t$, where $R^t$ represents the SU’s throughput. This negative correlation incentivizes the jammer to target channels where its actions most effectively degrade the SU’s performance. If the SU transmits on the same channel as the jammer, the resulting interference leads to a significant reduction in the achievable data rate. This adaptive, learning-based jamming strategy significantly outperforms random or fixed jamming approaches, presenting a more realistic and potent adversary in CIoT environments.

\section{Proposed Hierarchical DDPG Framework}
 The process of learning the optimal strategy that maximizes the sum rate of the CIoT network can be formulated as a Markov decision process (MDP). The MDP is represented by the tuple $(\boldsymbol{\mathcal{S}}, \boldsymbol{\mathcal{A}}, \boldsymbol{\mathcal{P}}, \boldsymbol{\mathcal{R}}, T)$, where $\boldsymbol{\mathcal{S}}$ signifies the set of CIoT environment states, $\boldsymbol{\mathcal{A}}$ represents the set of possible actions by the SU agent, $\boldsymbol{\mathcal{P}}$ represents the set of state transition probabilities, $\boldsymbol{\mathcal{R}}$ includes the rewards for specific state-action combinations, and $T$ indicates the time step. In the specified model-free MDP, the SU agent faces the challenge of assessing the value of state-action pairs without prior knowledge of $\boldsymbol{\mathcal{P}}$. However, through RL, the SU agent can approximate the state-value function and learn a policy $\pi$ that guides action selection $a_t$ per the current CIoT environment state $s_t$.

The state space of the MDP is defined as, $s_t = \left\{ B_n^t, e_n^{t-1}, \omega_{pm}^t, \omega_{jm}^t, g_{mn}^t, g_{nm}^t, g_{nn}^t \right\}$,
where $B_n^t$ is the current battery level, $e_n^{t-1}$ is the previously harvested energy, $\omega_{pm}^t$ and $\omega_{jm}^t$ are the PU and jammer activity indicators, respectively, and $g_{mn}^t$, $g_{nm}^t$, and $g_{nn}^t$ represent the channel gains. The overall action space is defined as, $a_t = [d_t, c^t, P_s^t]$,
where $d_t \in \mathcal{I} \triangleq \{0,1\}$ denotes the mode (harvest or transmit), $c^t \in \mathcal{C} \triangleq \{0, 1, \dots, M-1\}$ is the channel index, and $P_s^t \in \mathbb{R}$ is the continuous transmission power satisfying $0.01 \leq P_s^t \leq 0.1~\text{W}$. The reward $r^t$ is defined as the achievable rate in (\ref{eq:effective_rate}) if the agent adheres to all system constraints. Otherwise, a penalty of $-\phi$ is imposed to discourage constraint violations.

To effectively tackle the complex decision-making problem in CIoT networks, we extend DDPG into a three-level \textit{Hierarchical DDPG (H-DDPG)} structure that decomposes the action space into mode selection, channel selection, and transmit power selection. This hierarchical decomposition facilitates better learning and generalization in multi-dimensional continuous-discrete action spaces.

\textbf{1) High-Level Agent (Mode Selector):}  
The high-level agent operates as a discrete actor responsible for selecting the mode of operation at each time step, $ d_n^t \in \{0, 1\}$. The agent aims to maximize the expected return by choosing the optimal mode based on the current state $s_t$, as defined by:
\begin{equation}
    d_n^t = \arg\max_{d \in \{0,1\}} Q_H(s_t, d),
\end{equation}
where $Q_H$ is a Double Deep Q-Network (DDQN) that estimates the value of selecting mode $d$ given the state $s_t$. If the agent chooses to harvest energy (i.e., $d_n^t = 1$), no further action is required during that time slot. However, if the agent selects the transmission mode (i.e., $d_n^t = 0$), it must proceed to select both the transmission channel $c_n^t$ and the transmit power level $P_{sn}^t$.

\textbf{2) Medium-Level Agent (Channel Selector)}: the medium-level actor selects a channel $c_n^t \in \{0,1,\dots,M-1\}$ based on
\begin{equation}
    c_n^t = \arg\max_{c} Q_M(s_t, c),
\end{equation}
where $Q_M$ is a DDQN that estimates the value of selecting channel $c$.

\textbf{3) Low-Level Agent (Power Selector)}: The continuous actor $\mu_L(s_t)$ selects the transmit power $P_{sn}^t$ from a bounded range $[P_{\min}, P_{\max}]$. The power is trained using a DDPG actor-critic pair as in 
\begin{align}\small
    \mathcal{L}_{Q_L} &= \left( Q_L(s_t, P_{sn}^t) - \left[r_t + \gamma Q_L'(s_{t+1}, \mu_L'(s_{t+1})) \right] \right)^2, \\
    \nabla_{\theta^\mu_L} J &= \mathbb{E} \left[ \nabla_a Q_L(s, a) \nabla_{\theta^\mu_L} \mu_L(s) \right].
\end{align}
Importantly, the power selection at the low level is tightly coupled with the output of the mid-level agent, which determines the selected channel $c_n^t$. Specifically, the primary user's activity status $\omega_{pm}^t$, the interference constraint $I_{\text{th}}$, and the channel gains $g_{sp}^t$ and $g_{ss}^t$ are all functions of the selected channel. Therefore, the low-level agent's policy $\mu_L(s_t)$ must condition its output on the selected channel $c_n^t$, enabling it to choose a power level that satisfies interference constraints while maximizing throughput. The overall hierarchical action is thus defined as:
\begin{equation}
    a_n^t = 
    \begin{cases}
        \text{Harvest}, & \text{if } d_n^t = 1, \\
        (c_n^t, P_{sn}^t), & \text{if } d_n^t = 0.
    \end{cases}
\end{equation}

By decoupling the decision process into three levels, H-DDPG achieves more stable learning, improved exploration, and faster convergence in complex CIoT environments. 

\section{Simulation Model and Results}
In this section, we assess the effectiveness of our proposed H-DDPG strategy in enhancing the transmission efficiency of the EH-enabled CIoT network detailed in Section II. We consider a time-slotted transmission of $T=$30 slots, each with a duration $\tau$=1s. The primary network consists of $M=5$ PU Tx-Rx pairs that each use a maximum of $L$= 20 slots for transmissions using a power of $P_{pm}$= 0.2W. The interference threshold is $I_{th}$= 0.01W. The secondary network consists of an SU Tx-Rx pair. The SU is capable of EH, has a finite battery capacity of $B_{max}$= 0.5W, and the $E_{max}$=0.1W. Additionally, the SU can dynamically choose a transmit power \( P_{sn} \in \mathbb{R},\ 0.01 \leq P_{sn} \leq 0.1~\text{W} \). The initial harvested energy is $e_0$= 0 and the initial battery level is $B_0$= $B_{max}/4$. The channel power gains $g_{nn}^t$ and $g_{nm}^t$ follow an exponential distribution mean of 0.1 and 0.2, respectively. Without loss of generality, we consider $g_{nm}^t= g_{mn}^t$. For our proposed H-DDPG architecture, we utilize a hidden dimension of $64$ for all agents and a learning rate of $lr = 45 \times 10^-5$. The total number of training episodes is 2500, and we employ a mini-batch size of 140. A penalty value of $\phi=7$ is assigned when the SU agent violates the constraints outlined in (\ref{eqn:optim}) during the training process of our proposed DRL strategy. The replay buffer size is set at $\kappa=333$ experiences, and the discount factor is $\gamma=0.99$. 

To assess the effectiveness of the proposed H-DDPG framework for joint mode selection, power control, and channel access, we conduct a comparative evaluation against several baseline strategies: 1) Informed Upper Bound (IUB): This benchmark represents an idealized agent with perfect knowledge of the environment’s state. Although unrealistic in practice, it serves as a theoretical upper bound to gauge the maximum achievable performance. 2) Hierarchical DDQN (H-DDQN): In this variant, power control is treated as a discrete action, and the continuous-action DDPG components of H-DDPG are replaced with DDQNs. This comparison highlights the importance of modeling the hybrid (continuous + discrete) nature of the action space. 3) Flat DDPG (F-DDPG): This approach disregards the hierarchical structure of the action space and treats the entire decision-making process as a flat continuous-action problem. It serves to illustrate the limitations of non-hierarchical designs in handling multi-level decision processes \cite{10224324}. 4) Flat DDQN (F-DDQN): A traditional approach that employs a single DDQN agent to make joint decisions, ignoring both the hierarchical and hybrid aspects of the problem. This strategy is commonly used in the literature and serves as a baseline for evaluating the benefits of structured decision decomposition \cite{Nadia_Jamming_IoTJ_2024, DDQN_jamm_channelSelection}.

\begin{figure*}[!h]
    \centering
    \includegraphics[width = 2\columnwidth]{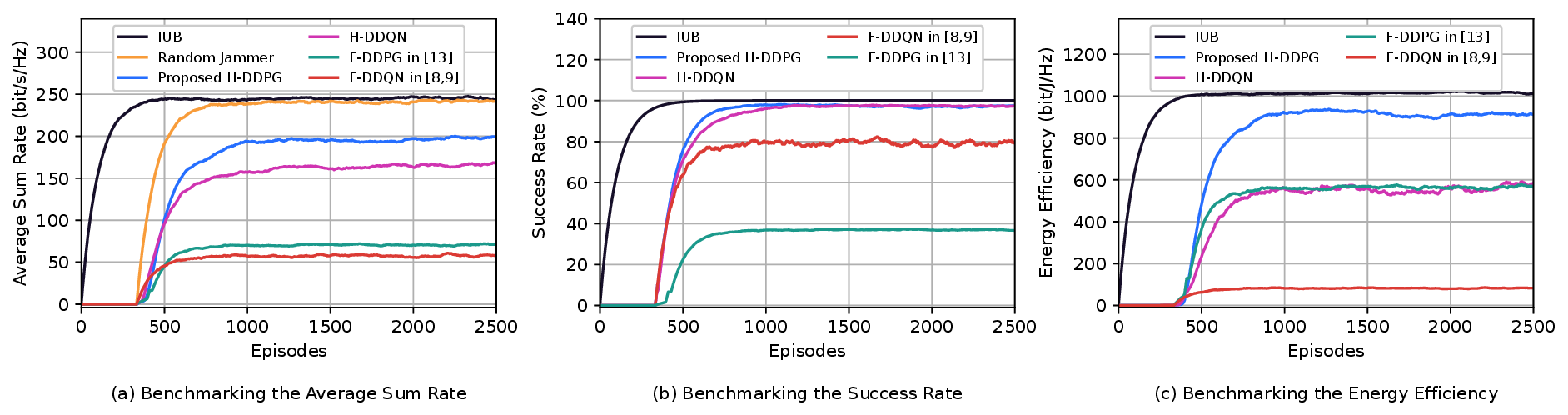}
    \caption{Benchmarking (a) the ASR performance, (b) the Successful transmission rate, and (c) the energy efficiency of our H-DDPG strategy in comparison to the existing strategies in the literature.}
    \label{fig:results}
\end{figure*}

In Fig.~\ref{fig:results}(a), we present a comparative analysis of the average sum rate (ASR) achieved by the SU using our proposed H-DDPG strategy alongside several benchmarks. The ASR reflects the moving average of the total sum rate defined in (\ref{eqn:optim}). As shown, all strategies converge within approximately 1000 episodes. The IUB represents the theoretical maximum ASR achievable in this CIoT environment, assuming the agent has perfect knowledge of the system state and can make optimal decisions accordingly. While the IUB consistently achieves the highest ASR, it is not implementable in practice due to its unrealistic assumption of full environmental observability. Thus, it serves as a reference upper bound. At the beginning of training, all learning-based agents exhibit near-zero ASR. This is expected, as they explore the action space with random decisions to populate the replay buffer with diverse experiences. Once sufficient experience is collected, the agents begin learning effective policies. 

As shown in Fig.~\ref{fig:results}(a), our proposed H-DDPG method consistently outperforms all other strategies and converges closest to the IUB benchmark, demonstrating its effectiveness in navigating the hybrid action space. The H-DDQN strategy surpasses both F-DDPG and F-DDQN, highlighting the benefit of hierarchical decision-making over flat structures. However, it still underperforms compared to H-DDPG, emphasizing the necessity of modeling the continuous nature of power control in addition to the discrete mode and channel selections. Among the flat architectures, F-DDPG outperforms F-DDQN, suggesting that continuous action modeling provides an advantage over discretized, monolithic control. Nonetheless, F-DDQN remains the least effective, reinforcing the importance of both hierarchical structuring and hybrid action modeling in complex CIoT environments. 
We also present a scenario, in Fig.~\ref{fig:results}(a), in which the proposed H-DDPG agent is evaluated against a \textit{random jammer} that selects a jamming channel uniformly at random in each time slot. The results demonstrate that the SU achieves a higher ASR in the presence of a random jammer compared to the smart jammer scenario. This highlights the significant impact of an intelligent adversary, a smart jammer, which adapts its actions based on the environment, poses a much greater threat to CIoT networks by strategically disrupting SU transmissions. The performance gap between the random and smart jammer cases underscores the importance of designing robust and adaptive access strategies, such as H-DDPG, to defend against adversarial interference in dynamic spectrum access environments.

In Fig.~\ref{fig:results}(b), we show a detailed comparison of the successful transmission rate achieved by the SU agent using our proposed H-DDPG strategy with other baseline strategies from the literature. The success rate serves as a measure of the agent's reliability under dynamic and adversarial conditions, and is defined as the ratio of successful transmissions to the total number of transmission attempts. A transmission is considered successful if the agent chooses to transmit and satisfies all the constraints outlined in the optimization problem (\ref{eqn:optim}). As expected, the IUB strategy achieves a 100\% success rate, reflecting the performance of an agent with perfect knowledge of the environment. Our proposed H-DDPG strategy closely matches the IUB, achieving a near-perfect success rate. Notably, it converges more quickly than the H-DDQN approach, highlighting the advantage of using a continuous-action algorithm in a hierarchical framework. Both hierarchical approaches (H-DDPG and H-DDQN) significantly outperform the flat strategies (F-DDPG and F-DDQN), underscoring the effectiveness of modeling the decision-making process hierarchically in complex CIoT environments. Interestingly, the F-DDQN strategy surpasses the F-DDPG in terms of success rate, suggesting that when using flat architectures, discrete-action algorithms like DDQN may be more effective than their continuous-action counterparts in learning to satisfy strict transmission constraints.

In Fig.~\ref{fig:results}(c), we present a comparative analysis of the energy efficiency (EE) achieved by the SU agent using our proposed H-DDPG framework against several benchmark strategies. EE is a vital performance metric in energy-constrained cognitive radio networks, defined as the ratio of the total achievable sum rate to the total energy consumed for transmission. This metric captures how effectively the SU converts its limited energy resources into successful data transmission. Crucially, this definition incorporates the SU’s battery dynamics and energy harvesting behavior, as the energy available for transmission is bounded by the current battery level and shaped by prior harvesting decisions. 

As shown in Fig.~\ref{fig:results}(c), the IUB, which assumes perfect knowledge of the environment, achieves the highest energy efficiency. Our H-DDPG framework outperforms all others, closely approaching the performance of the IUB. This highlights the algorithm’s ability to balance transmission and harvesting decisions while adapting to fluctuating energy availability and environmental uncertainty. The H-DDQN and F-DDPG strategies exhibit similar EE performance, both surpassing F-DDQN, but neither reaching the efficiency of H-DDPG. This emphasizes the importance of modeling the hybrid nature of the action space: while H-DDQN benefits from a hierarchical structure, its discretized power levels reduce flexibility; conversely, F-DDPG supports continous action representation but lacks a hierarchical structure to effectively coordinate decisions across different levels. Finally, the F-DDQN strategy records the lowest EE, reinforcing the limitations of ignoring both the hybrid and hierarchical aspects of the decision-making process in dynamic spectrum environments.

\begin{figure}[!t]
\centering
\includegraphics[width=0.8\linewidth]{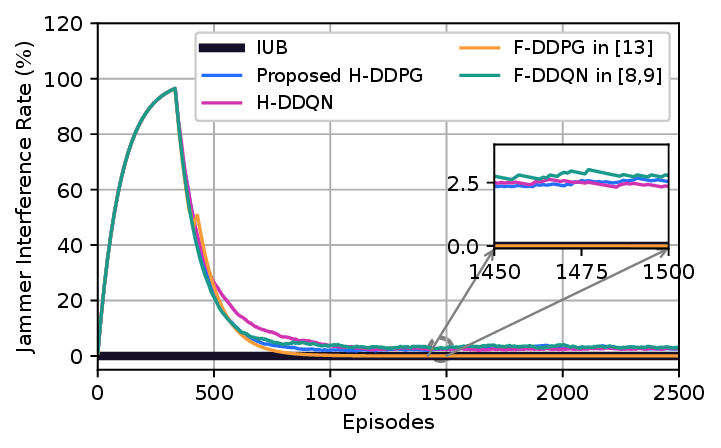}
\caption{Benchmarking the jammer interference rate of our H-DDPG strategy in comparison to the existing strategies in the literature.}
\label{fig:jamm}
\end{figure}

In Fig.~\ref{fig:jamm}, we evaluate the jammer interference rate, which measures the jammer’s ability to disrupt SU transmissions. This metric is defined as the ratio of the number of SU transmission attempts that occur on the same channel as the jammer’s selected channel, to the total number of SU transmission attempts. It reflects both the jammer's effectiveness and the SU agent’s ability to avoid interference over time. As shown in the figure, the IUB strategy achieves a jammer interference rate of zero across all episodes, thanks to its perfect knowledge of the environment.  For all learning-based strategies, the interference rate initially rises close to 100\% during the exploration phase, where agents select actions randomly to fill their replay buffers. Once sufficient experience has been collected and learning begins, the interference rate gradually decreases and stabilizes after approximately 1000 episodes. Notably, the F-DDPG strategy achieves the lowest interference rate among learning-based methods, closely approaching the IUB. The remaining strategies, including our proposed H-DDPG, converge to an interference rate of around 2.5\%. Although the H-DDPG does not eliminate interference entirely, this small residual rate is negligible when considering the overall system performance. Our H-DDPG strategy still achieves the highest ASR, success rate, and EE.

\begin{figure}[!t]
\centering
\includegraphics[width=0.85\linewidth]{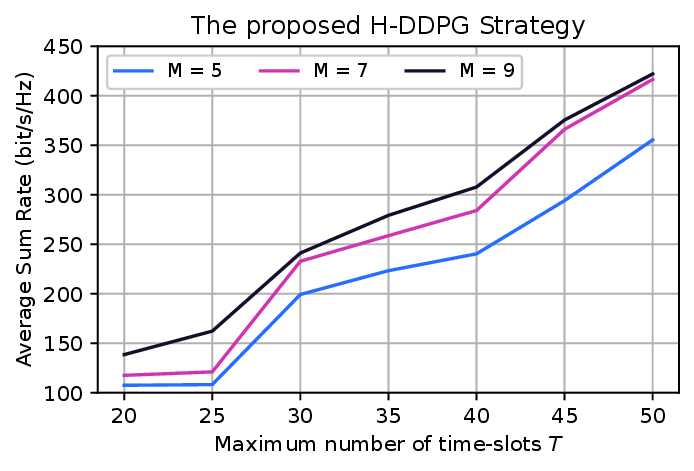}
\caption{The effects of varying the maximum number of time slots $T$ and the number of channels $M$ on the ASR of our proposed H-DDPG strategy}
\label{fig:T_M_Bmax}
\end{figure}

In Fig.~\ref{fig:T_M_Bmax}, we investigate the impact of varying the number of time slots $T$ and the number of available channels $M$ on the ASR performance of the proposed H-DDPG strategy. As illustrated in the figure, increasing the time horizon $T$ leads to an improvement in ASR. This is because a longer time horizon provides more opportunities for the SU agent to identify idle slots, thus reducing potential interference with PUs and enabling the selection of higher transmission powers more frequently. Similarly, an increase in the number of available channels $M$ results in better ASR performance. A larger channel set offers the SU agent greater flexibility in avoiding both PU activity and jamming interference. With more channel options, the probability of finding an idle and unjammed channel during any given time slot increases, thereby enhancing the SU's ability to transmit with higher efficiency and throughput. 

\section{Conclusion}
In this paper, we proposed a novel H-DDPG framework for joint mode selection, power control, and channel access in energy-constrained CIoT networks operating under the presence of a smart jammer. Our approach decomposes the complex decision-making process into hierarchical levels to address the challenge of mixed discrete-continuous action spaces. The SU agent learns to dynamically choose between transmission and energy harvesting, select the optimal channel, and adjust transmission power, all while satisfying interference and energy constraints. Concurrently, the smart jammer is modeled as an RL agent that adaptively learns to disrupt SU transmissions by selecting optimal jamming channels. Extensive simulation results confirm that the proposed framework significantly improves sum rate, success rate, and energy efficiency, while maintaining strong resilience to jamming, consistently outperforming baseline methods and closely approaching the ideal upper bound.

\bibliography{ref.bib}
\bibliographystyle{IEEEtran}
\end{document}